# Temperature and Pressure-Induced Atomic Structure Evolution During Solidification of $Zr_{50}Nb_{50}$ Metallic Melt via Molecular Dynamics Simulation

Frew Gashaw Asefa[1], Yi Ma[2], Qin Wu[3], Yedilfana Setarge Mekonnen[4,*], Tekalign Terfa Debela[5,*]

[1]Material Science Program, College of Natural and Computational Sciences, Addis Ababa University, P. Box 1176, Addis Ababa, Ethiopia

[2]Key Laboratory for Light-weight Materials, Nanjing Tech University, Nanjing 210009, PR China

[3]Center for Functional Nanomaterials, Brookhaven National Laboratory, Upton, NY 11973-5000 USA

[4]Center for Environmental Science, College of Natural & Computational Sciences, Addis Ababa University, Ethiopia

[5]Institute for Application of Advanced Materials, Jeonju University, Chonju, Korea

*Corresponding authors: yedilfana.setarge@aau.edu.et, and debelatt@gmail.com

## Abstract

In this report, the evolution of the local atomic structure of the $Zr_{50}Nb_{50}$ melt was investigated by applying temperature (2600 to 300 K) and pressure (0 to 50 Gpa) using classical molecular dynamics simulations. To gain clear insight into the structural evolution during quenching, we used various methods of structural analysis such as the radial distribution function g(r), coordination number, bond angle distribution, and Voronoi tessellation. We found that the icosahedral motifs (which are the signature of the short-range ordering) and distorted BCC-like clusters dominate in the liquid and glass region under 0 and 5 Gpa external pressure. A first-order phase transition to a crystal-like structure was observed at 10, 15, and 20 Gpa external pressure at 1400, 1500, and 1600 K, respectively. Before the first-order phase transition, the system was dominated by icosahedral and distorted BCC-like clusters. When the temperature is lowered further below the glass transition at 10,15, and 20 Gpa external pressure, all structural analyses show that the solidified system consists mainly of body-centered cubic-like clusters in the case of our specific cooling rate of $10^{12}$ K/s.

# 1. Introduction

The study of metallic melts and metallic glasses is challenging because the atoms are randomly distributed in space, which means that there is no long-range order, but they are populated by short- and medium-range order [1-5]. Moreover, the short- and medium-range orders are interconnected in metallic melts and metallic glasses. Depending on the type of interconnection, they are populated by different atomic clusters [1,3-6]. It is known that the evolution of clusters in metallic glasses is affected by external pressure and temperature. However, it is unclear how the clusters develop as a function of external influences (e.g., pressure and temperature). Therefore, the study of metallic liquids is fundamental to understanding the melting, solidification, and glass transition processes, and plays a key role in deciphering the chemical order that gives metallic glasses unique physical properties compared to their crystalline counterparts, such as high strength [7], high toughness [8], large elastic limit [9], high corrosion [10], and wear resistance [11].

High-energy X-ray and neutron diffraction are the best known and most common experimental methods for structure determination in metallic melts [12,13]. Classical molecular dynamics and *ab initio* molecular dynamics simulations are the most commonly used simulation techniques for alloy structure determination [14-16]. Characterizing the evolution of atomic structure during solidification or phase transition is challenging with the above experimental techniques, especially because metallic glasses form when the melt is quenched at high cooling rates ($10^3$-$10^{12}$ K/s) to avoid crystallization [17,18]. Moreover, it is not possible with current experiments to directly measure how an atom in a crystal lattice is transported to a particular neighbor and which local structures tend to rearrange. For these reasons, it is quite difficult or nearly impossible to determine experimentally how the applied pressure affects the transition from a liquid to a solid. Computer simulations and molecular dynamics methods can be used to overcome the difficulties in experimental methods for characterizing alloy structures at high cooling rates.

Zr-Nb alloys are structural materials usually selected for their good mechanical and corrosion-resistant properties [10,19]. To date, many researchers have conducted numerous studies on Zr-Nb alloys, both theoretically [20-23] and experimentally [10,25,26]. However, the effects of pressure on the evolution of the atomic structure of the Zr-Nb system at different compositions and temperatures have hardly been studied. Apart from the fact that hydrostatic pressure is another variable that can be used to tune material properties, the effects of pressure on the structure of molten metals have not been thoroughly studied. Yang *et. al.* [20] reported that $Zr_{50}Nb_{50}$ exhibits a glass structure at lower temperatures.

The study of the effects of pressure and temperature on the structure is very important to optimize the properties. Therefore, in this study, the effects of pressure and temperature on the local atomic structure of $Zr_{50}Nb_{50}$ were investigated by applying atomic analysis such as radial distribution function, coordination number, bond angle distribution, and Voronoi tessellation. The changes in local atomic structure were analyzed by quenching from 4000 – 300 K with a cooling rate of $10^{12}$K/s, and a temperature interval of 200 K. The effects of pressure on the local atomic structure were determined by subjecting the $Zr_{50}Nb_{50}$ composition to different pressures (0, 5, 10, 15, and 20 Gpa). Separate calculations were performed for each pressure, keeping the quenching temperature and cooling rate the same.

## 2. Simulation Details

We have performed a classical molecular dynamics (MD) simulation for $Zr_{50}Nb_{50}$ to study the evolution of the atomic structure during solidification from 2600 to 300 K at a cooling rate of $10^{12}$ K/s. A cubic box with 6400 randomly distributed Zr and 6400 Nb atoms were used to simulate the $Zr_{50}Nb_{50}$ system. The calculations were performed using the open-source parallel simulator LAMMPS. The simulation cell was subjected to periodic boundary conditions (PBC) along the x, y, and z axes to avoid surface formation. A molecular dynamics time step of 2 fs was set to integrate the Newtonian equation of motion using the velocity verlet algorithm as implemented in LAMMPS. The system was melted to 4000 K in the isobaric isothermal ensemble (NPT) and relaxed for 200 ps. No pressure is applied during melting. The melt was quenched to 300 K under 0, 5, 10, 15 and 20 Gpa external pressure using the NPT ensemble with a 200 K temperature interval and all data for statistical characterization were extracted from a system equilibrated for 100 ps at each temperature. The atomic interaction was modeled by angular dependent potential developed by Smirnova for the Zr-Nb system [21]. The total energy of the system is given in equation (1) below.

$$U = \sum_{i>j} \varphi_{\alpha\beta}(r_{ij}) + \sum_i F_\alpha(\rho_i) + \frac{1}{2}\sum_{i,k}\left(\mu_i^k\right)^2 + \frac{1}{2}\sum_{i,k,l}\left(\lambda_i^{kl}\right)^2 - \frac{1}{6}\sum_i V_i^2 \quad (1)$$

Here the indices $i$ and $j$ enumerate the atoms, while the superscripts $k$, $l$ = 1, 2, 3 refer to the Cartesian components of vectors and tensors. The subscripts $\alpha$ and $\beta$ denote the element types of the atoms. The first term in equation (1) represents the pair interactions between atoms via a pair potential U. The summation is over all $j^{th}$ neighbors of the $i^{th}$ atom within the cut-off distance $r_{cut}$ 6.2 Å. The second term F is the embedding energy, which is a function of the total electron density $\rho$. The first two terms in Eq. (1) provide a main contribution to the energy of the system. Additional $l$ and $k$ terms introduce non-central interactions through the dipole vectors and quadrupole tensors. They are intended to penalize deviations of the local environment from the cubic symmetry.

## 3. Results and Discussion

### 3.1 Energy-temperature (*E-T*) and volume-temperature (*V-T*) curves analysis

Figure 1 (a) shows the change in volume as a function of temperature when the system is cooled under the external pressure of 0, 5, 10, 15, and 20 Gpa. The result shows that the phase transition temperature point shifts to a high higher temperature in line with high-pressure values. This scenario is also reported in other simulation works  For the system under 0 and 5 Gpa external pressure, the volume and energy (see Figure 1 (b), decreases linearly, and no abrupt change is observed which means a glassy state is retained in the rapid solidification process. A sudden jump to lower values is observed on both *V-T* and *E-T* curves on the application of external pressure values of 10, 15, and 20 Gpa. For 10 Gpa pressure as temperature drops from 2600 to 1800 K, the energy decreases gradually. However, the sudden jump at 1600 K shows that the system undergoes first order (liquid to crystalline) phase transition. For 15 Gpa and 20 Gpa sudden jump is observed at higher temperature values of 1500 and 1600 K respectively. At 2600 K, the arrangement of atoms is disordered. This means that the system is in the liquid (amorphous) state, where the atoms are distributed over the space randomly. We want to recall that an amorphous (non-crystalline) material is solid without long-range ordering. Thus, we can only characterize within short-range ordering  To achieve this, we use various structural analysis methods in the following sections.

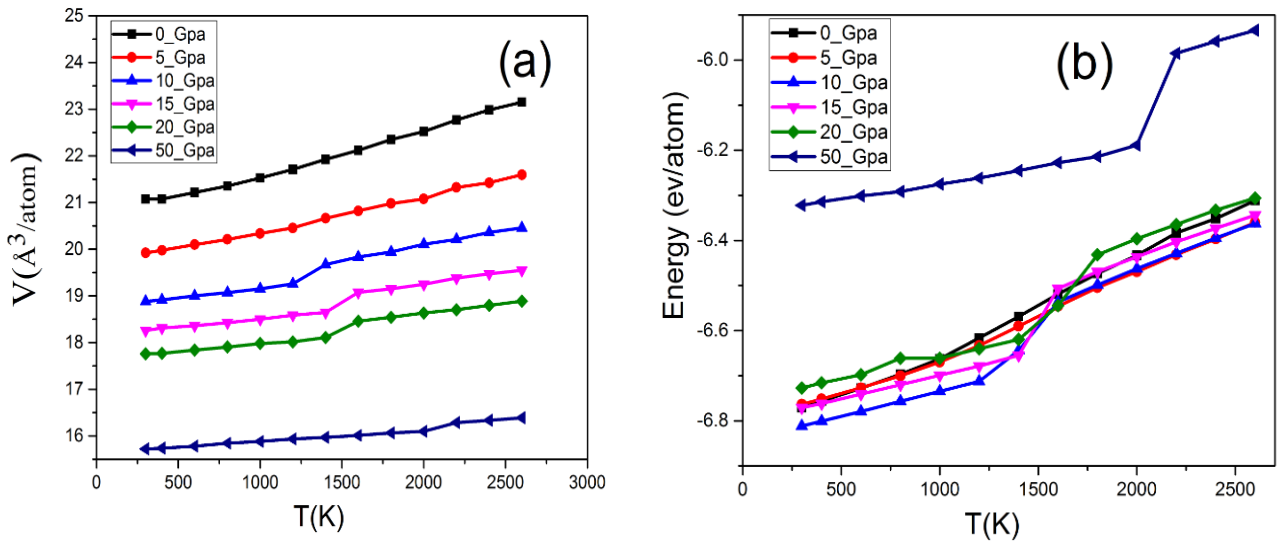

**Figure 1**. (a), Volume, (b) internal energy per atom of $Zr_{50}Nb_{50}$, versus temperature at a cooling rate of $10^{12}$K/s

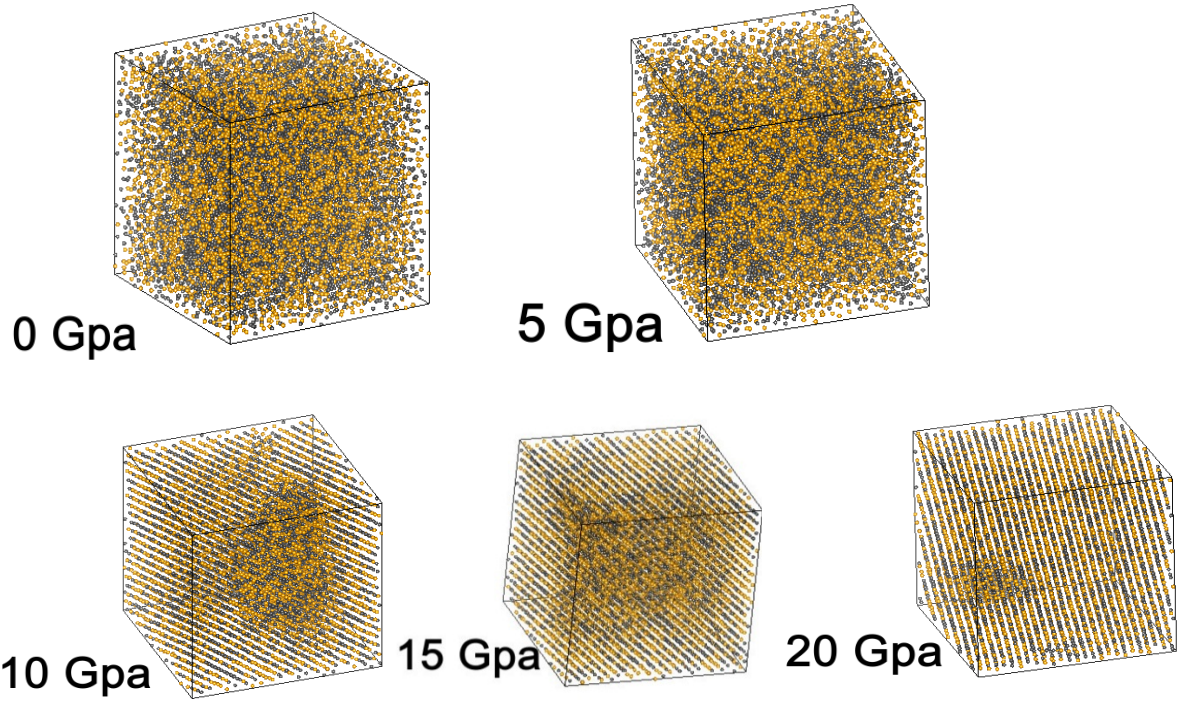

**Figure 2.** Simulation snapshot of atomic distribution at 300 K under external pressure of 0, 5, 10,15, and 20 Gpa, Zr atoms are represented by yellow color and Nb atoms are represented by black color.

## 3.2 Pair distribution function (PDF, g(r))

In the study of metallic liquids systems or amorphous systems, the pair correlation function is used to predict the static structures and more importantly, it is used to describe the statistics of each atomic species around the other species in a system. It gives the probability that an atom is at distance *r* from the reference atom and can be calculated as follows:

$$g(r) = \frac{1}{4\pi r^2 \rho_o N} \sum_{i \neq j} \sigma(r - r_{ij}) \qquad (2)$$

where $\rho_0$ is the average number density of the system, $N$ is the total number of atoms, and $r_{ij}$ is the interatomic distance between atoms $i$ and $j$. The partial pair correlation functions for the constituents in our alloy is calculated by;

$$g_{a-b}(r) = \frac{N}{4\pi r^2 \rho_o N_a N_b} \sum_{i=1}^{N_a} \sum_{j=1}^{N_b} \sigma(r - r_{ij}) \qquad (3)$$

where $N_a$ and $N_b$ are the numbers of atom types a and b, respectively.

The total PDF and partial PDF curves for all pressures of 0, 5, 10, 15, and 20 Gpa were plotted in the temperature ranges of 2600 – 300 K. Figure 3 depicts the total and partial PDF at 0 Gpa external pressure. The total pair distribution function (see Figure 3 (a)) has a typical shape with two main peaks and broad peaks. The intensities of these peaks, increase as temperature decreases. This is a clear indication that $Zr_{50}Nb_{50}$ melts are achieving a sort of local ordering. At 800 K the possibility for the second peak splitting is observed. At 600 K, the second peak splits into two sub-peaks at r = 5.24 and 5.8 Å. On further cooling down to 300 K, the split peak intensity enhances, and the first pre-peaks shift to lower r = 5.08 Å, and the second pre-peaks shift to higher r = 5.96 Å. This is the signature for short-range ordering (amorphization) in the system within this temperature environment. The contribution of each species to this effect can be well understood from plots of the partial pair distribution function.

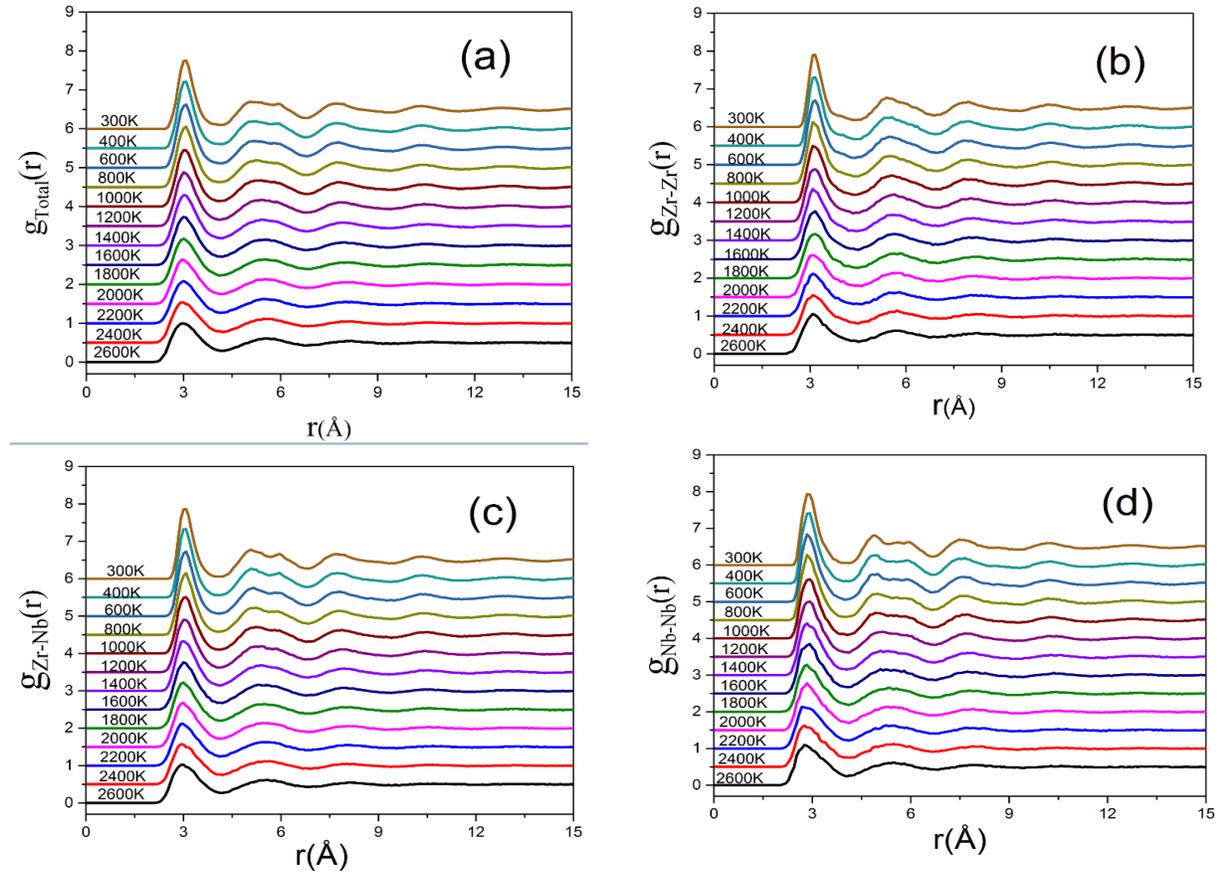

**Figure 3**. (a) Total PDFs for $Zr_{50}Nb_{50}$ under 0 Gpa pressure, from 2600 – 300 K, (b), (c), (d) PPCFs for Zr-Zr, Zr-Nb and Nb-Nb from 2600 –300 K, respectively.

In Figure 3 (d), the Nb-Nb pair distribution function shows no full second peak splitting in the temperature range of 2600 – 300 K, rather exhibits the evolution of sub-peak followed by a broad plateau starting from 800 – 300 K. This indicates Nb – Nb has no contribution to the second peak splitting which is the characteristics feature of metallic glass. Also, the Zr-Zr pair doesn't show second peak splitting in all temperature ranges as shown in Figure 3 (b). The Zr-Nb pairs (See Figure 3 (c)), show the second peak splitting at 1200 K, as the system further cooled down to 300 K, the peak becomes more enhanced. Therefore, Zr-Zr and Nb-Nb pairs do not contribute to the splitting in $Zr_{50}Nb_{50}$ binary alloy TPDF. This result was also observed in other theoretical work[26]. Thus we can conclude that the amorphous (metallic glass) structure formed at 300 K is mainly initiated by the Zr-Nb interaction. Moreover, we analyzed each species' contribution to total PDF at 300 K in terms of interatomic distance from the PPCF at 300 K and their sum of Goldschmidt radius. Goldschmidt ($R_{Zr}$ = 1.60A and $R_{Nb}$ = 1.46A). The first peak position of the total PDF is located at **r** = 3 Å. The first peak of Zr-Zr, Nb-Nb, and Zr-Nb is located at 3.16, 2.84, and 3 Å respectively. Besides their sum of Goldschmidt radii is 3.2, 2.92, and 3.04 Å, respectively [27]. The first peak position of Nb-Nb and Zr-Nb is smaller than their corresponding interatomic distance.

The total and partial PDF for the system under 5Gpa, external pressure was shown in Figure 5. As we cool down from 2600 – 300 K the intensities of the first and second main peaks increase. The difference in the first peak intensity, position, and second peak shape is observed, compared with the system under 0 Gpa pressure. The first and second peak intensity increase from 2600 – 1400 K, however on further cooling down from 1300 – 300 K the second peak intensity decreases and it became flattened to a broad plateau. More importantly, unlike at 0 Gpa pressure splitting in the second peak is observed at only 300 K, indicating the presence of numerous short-range orderings at 300 K. This indicates the overall local structure variation on the application of 5 Gpa external pressure. The numerical values of peak positions are given in Supporting Information Table S1.

As one can see from Table 1, the first peak position of total PDF is located at r = 2.92 Å from 2600 – 2000 K, and it remains at r =3 Å, from 1800 – 300 K. At 5 Gpa pressure the peak position shows a constant value of 2.92 Å, from 2600 – 2000 K, and it shifts to 3 Å on cooling down from 1800 – 300 K. The first peak position of TPDF shifts from its value 3 Å under 0 Gpa pressure to 2.92 Å on the application of 5 Gpa pressure. This is a clear indication of some pressure-related changes. From the partial pair distribution function see Figure 4. (b), (c), (d) the first peak position of Zr-Zr is located at r = 3.08 Å, form 2600 – 2200 K and it shifts to r = 3 Å, from 1800 – 1200 K. Finally, it shifts two 3 .08 Å form 1000 – 300 K. Also, the second peak position shifts to smaller r = 3.32 Å at 300 K. At higher temperature, the second peak position fluctuates between r = 5.4 and r = 5.64 Å. It attains a constant value of 3.32 Å, from 800 – 300 K and finally, this second peak is split into two sub-peaks located at r = 5.32 and r = 5.72 Å. The Nb-Nb pair's first peak position fluctuates between r = 2.8 and 2.76 Å. From 1200 – 300 K, it shifts to a constant value of 2.84 Å. at 1000 K, the second peak splits into two sub-peaks, where the first sub-peaks located at r = 4.84 Å and the second sub-peak located at r = 5.4 Å. On further quenching to 300 K, the sub-peak intensity increases while holding an almost constant value of peak position (r = 4.84 Å). The second sub-peaks slightly shift to the right at 300 K. Compared with the Nb-Nb pairs at 0 Gpa pressure, Nb-Nb pairs show a similar first peak position at 300 K. More importantly unlike at 0 Gpa, the Nb-Nb pairs show second peak splitting starting at 1000 K.

The first peak position of Zr – Nb varies by 0.8 Å in quenching from 2600 – 1800 K. It maintains a constant value of 3 Å between 1600 and 300 K. While the second peak, position fluctuates between 5.24 at 1400 K and 5.48 at 2600 K. The second peak attains large peak position value at high temperature (See Supporting Table 1.). As one can see from the table the shifts in the second peak position are by a factor of 0.08A, in a temperature range of 2600 – 1400 K. From 1400 -1200 K, it shifts to the right by 0.04 Å. From 1000 – 600 K, it remains at a constant value of 5 Å. At 600 K the second peaks split into two sub-peaks positioned at r = 5 and 5.96 Å. These peaks were found at r = 5.08 Å and 5.96 Å at 300 K with enhanced intensity. Compared with structural evolution at 0 Gpa, Zr – Nb, Nb –Nb pairs shifts to similar first peak position value. A difference is observed in the Zr-Zr pairs' first peak position, it shifted to a smaller r-value at 5 Gpa external pressure than 0 Gpa. The shifts of TPDF's first peak position at 5 Gpa to 3 Å from, its value of 3.08 Å at 0 Gpa are caused by the behavior of Zr – Zr bonding sensitivity to pressure. This behavior of Zr is also observed in other previous works by Sengul et al. [28].

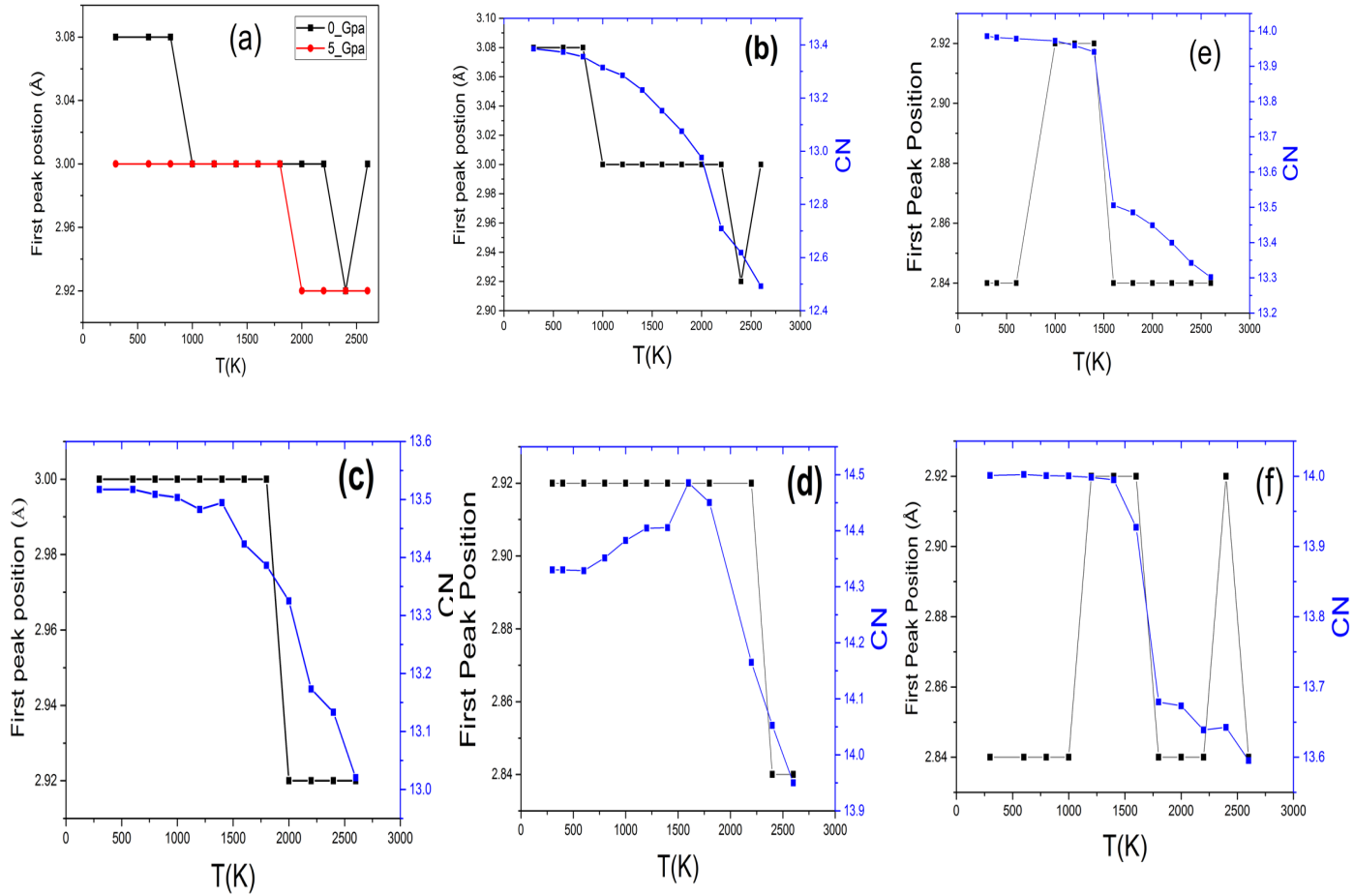

**Figure 4.** (a), the first peak position of g(r) with a decrease of temperature under 0 and 5 Gpa external pressure, (b-f) are the first peak position of total g(r) and CN, evolution at each temperature at 0, 5, 10, 15, and 20 Gpa external pressure respectively.

To see changes in the first peak position, of total PDF on quenching from 2600 – 300 K for the same systems under 0 and 5 Gpa external pressure, we have plotted the first peak positions of total PDF versus temperature in Figure 4, (a). As one can see from the figure, at 300 K, the peak position is shifted to 3 Å at 5 Gpa external pressure, from its value of 3.08 Å under 0 Gpa external pressure. From the overall trend of peak position for the two pressure values, the peak positions shift to larger r values, on quenching from 2600 – 300 K. This indicates the expansion of local structures in the first coordination shells. As we increase the pressure values to 5 Gpa the reduction of the first peak position observed indicates, that pressure hinders the expansion of the local structure while increasing the local packing efficiency. As the pressure is increased, these phenomena become more apparent (see Fig.4(b-f)). To quantitatively reveal this phenomenon, we have calculated the coordination number, by integrating the area under the first peak of the total RDFs up to the first minimum position of its first peak. The plots are shown in Figure 4. insets (b-e). At 0 Gpa external pressure as temperature drops from 2600 – 2400 K, the peak position, decreases from 3 to 2.92 Å and returns to 3 Å as temperature drops from 2400 – 2200 K and maintains a constant value of 3 Å up to 1000 K and finally increases to 3.08 Å and maintains this value, from 1000 – 300 K. While the CN increases smoothly from 12.5 at 2600 K to 12.71 at 2200 K. It increases to 13.4, as temperature drops to 1800 K.  As temperature further cooled down to 300 K, the CN number smoothly increases to a maximum value of 13.4, Indicating the local structure of bcc and amorphous.

At 5 Gpa external pressure, Figure 4 (c), the peak position maintains a constant value of 2.92 Å as temperature drops from 2600 – 2000 K and at 1800 K it increases to 3 Å, and maintains 3 Å from 1800 – 300 K. While the CN increases smoothly from 13.02 at 2600 K, to 13.173 at 2200 K, followed by a large increase to 13.325 at 1800 K.  Again, smooth increase to 13.5, accompanied by small fluctuations, is observed as temperature further decreases from 1800 - 1200 K followed by small decreases at 1000 K to 13.482. Finally, the CN increases to 13.517 at 300 K.  Indicating the local structure of amorphous at high and low temperatures. At 10, 15, and 20 Gpa (see Fig.4(d-f)), external pressure the coordination number shows the local structure of amorphous above 1400, 1500, and 1600 K respectively, and bcc crystal structure below respective temperature.

The pair distribution function under 10 Gpa external pressure is shown in Figure 6.  As one can see from the Figure 6 (a), the total pair distribution function shows a clearly smooth and broader shape trend in the temperature range of 2600 – 1800 K. A second peak is shifted to smaller r values along with the evolution of pre-peak at 1600 K and the pre-peak grows slowly at becoming obvious shoulder peak at r = 4.85 Å.  Zr-Zr and Zr-Nb pairs possess similar peak positions with their peak positions under 5 Gpa, which is 3 and 2.92 Å respectively. The Nb-Nb pairs position is reduced to r = 2.76 Å from its value of 2.84 Å at 5 Gpa. At 1400 K a sudden change in the shape of total PDFs and PPDFs is detected. The curve loses the whole pattern it possesses above 1400 K. The first peak becomes sharpened with increased intensity. The pre-peaks observed at 1600 and 1500 K on the second peaks become increased to splitting the second peaks into two peaks.

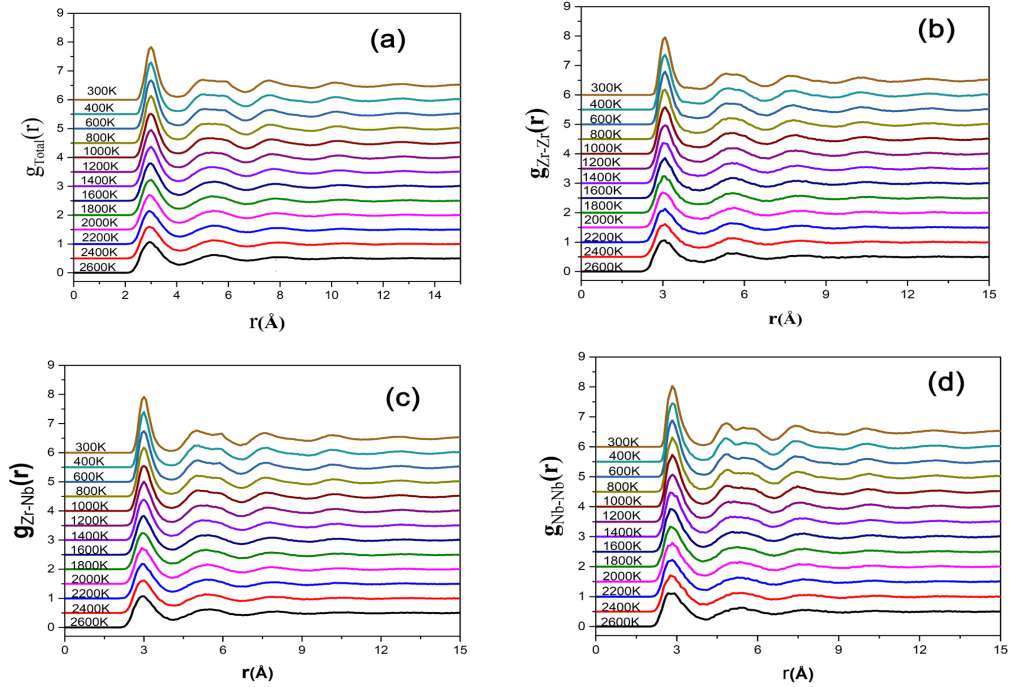

**Figure 5.** (a) Total PDFs for $Zr_{50}Nb_{50}$ under 5 Gpa pressure, from 2600 – 300 K, (b), (c), (d) PPDFs for Zr-Zr, Zr-Nb and Nb-Nb from 2600 – 300 K, respectively.

The obvious evolution of the third, fourth, fifth, and sixth sharp peaks is observed. The intensities of these peaks become more pronounced on further cooling down to 300 K. The first peak shows the evolution of the shoulder peak to the right, at 600 K, and becomes more pronounced on cooling down to 300 K. Here the scenario is as we cool down the system from 2600 – 1500 K, the width of the first and second peaks becomes narrowed smoothly along with the emergence of pre-peaks on the second peak at 1600 and 1500 K. This indicates the presences of SRO or MRO in the system. The sudden change observed below 1400 K indicates the supercooled liquid, undergoes a first-order phase transition to a crystalline structure. The energy and volume jump observed below 1400 K, as shown in Figures 1 (a) and (b) crystallization occurred below 1400 K.

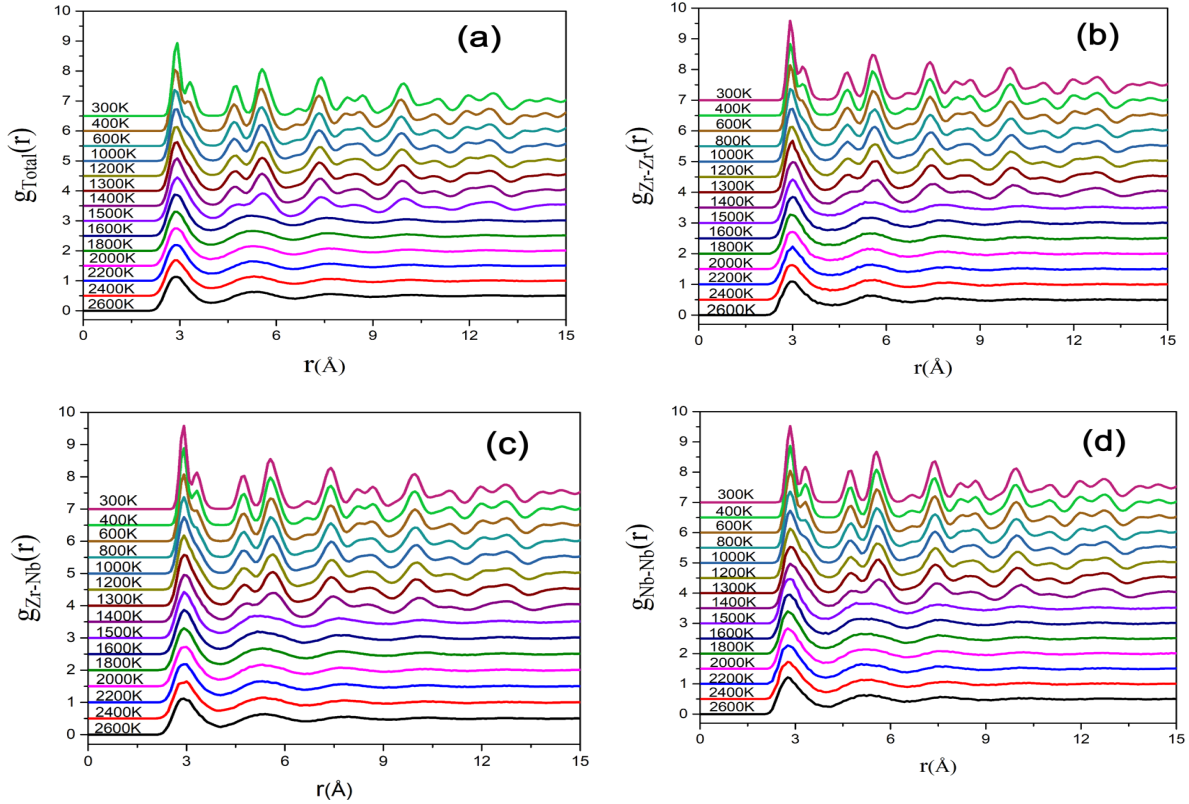

**Figure 6.** (a) Total PDFs for $Zr_{50}Nb_{50}$ under 10 Gpa pressure, from 2600 – 300 K, (b-d) PPDFs for Zr-Zr, Zr-Nb and Nb-Nb from 2600 – 300 K, respectively.

The pair distribution function under 15 Gpa is shown in Figure 7. As one can see from Figure 7. (a), the total pair distribution function, shows broad first and second peak peaks located at r = 2.84 Å and r = 5.4 Å at 2600 K respectively. The first peak position fluctuates, between 2.84 and 2.92 Å on cooling down to 1600 K and maintains a constant value of 2.92 Å from 1500 – 1000 K and shifts to r = 2.84 Å on further cooling down to 300 K. while the second peak is located at r = 5.4 Å at 2600 K. As we cool down from 2600 – 1600 K, the second peak maintains a constant value of 5.24 Å. At 1500 K the TPDF second peak reduces in intensity and split into two peaks located at r = 4.84 and r = 5.56 Å. On further cooling down to 300 K, the pre-peak enhances in intensity, and its position fluctuates between 4.84 and 4.68 Å while the second main peaks maintain a constant value of 5.56 Å. This indicates crystallization occurs below 1500 K, at 15 Gpa external pressure. The calculated partial pair distribution function is also given in Figure 7. insets, (b), (c), and (d). The partial pair distribution function for Zr-Zr, see Figure 7. insets (b), the first peak position, is located at r = 3 Å from, 2600 – 1800 K, while the second peak fluctuates between 5.4 and 5.56 Å. The first peak position reduces to r = 2.92 Å at 1600 K and maintains a constant value of 2.92 Å on further cooling down from 1600 – 300 K. While the second peak split into two peaks located at r = 4.76 and 5.56 Å. On further cooling down to 300 K, the peak intensities, become enhanced and shift to lower r = 4.68 A and 5.46 A respectively.

Zr-Nb pairs see Figure 7. (c), which shows constant first peak maxima at r = 2.92 Å from, 2600 – 1000 K while the second peak position fluctuates between 5.24 and 5.08 Å. The second peaks split into two peaks located at r = 4.84 and 5.56 Å respectively at 1500 K. The first peak position shifts to smaller r = 2.84 Å on further cooling down from 1000 – 300 K. Possibility of shoulder peak evolution on the first peak is observed at 800 K, this shoulder peak becomes obvious at 400 K. On cooling from 400 – 300 K the first main peak shifts to r = 3.32 Å.

Nb-Nb, pairs, see Figure 7. (d), the first peak position is located at r = 2.68 Å, it increases to r = 2.84 Å, as we cool from 2600 – 2400 K. From 2200 – 1600 K, the first peak position maintains, a constant value of 2.76 Å. As we further cooled down from 1500 – 300 K, the first peak position shifted to a higher r-value of 2.84 Å. While the second peak is located at r = 5.16 Å at 2600 K and 2400 K. From 2200 – 1800 K, it shifts to r = 5.08 Å and it increases to 5.24 Å at 1600 K. At 1500 K the second peak splits into two peaks (pre-peak and main peak) located at r = 4.76 and 5.56 Å. On further cooling down from 1400 – 300 K, the height of pre- peaks becomes pronounced and maintains a constant, peak position of r = 4.68 Å while the second main peak first holds a constant value of 5.56 Å from, 1500 – 1000 K, and then after shifts to r = 5.48 Å on cooling from 800 – 300 K.

The pair distribution function for 20 Gpa is shown in the Figure.8, As one can see from Figure.8, (a) the total pair distribution function, shows broad first and second peak peaks located at r = 2.84 Å and r = 5.16 Å at 2600 K respectively. As we cool down from 2600 – 1600 K the first peak position fluctuates between 2.84 and 2.92 Å. On further cooling down from 1500 - 300 K, the first peak position shifts to r = 2.84 Å. At 1000 K, the first peak becomes sharpened and evolution of the shoulder peak at r = 3.24 Å is observed at 600 K. The height of this shoulder peak becomes increases on cooling down to 300 K while maintaining a constant peak position of 3.24 Å. The second peak also shows fluctuation by a constant value and splitting. First, the second peak is located at 5.16 Å at 2600 and 2400 K, but it increases to 5.32 Å at 2200 K. From 2000 – 1700 K the second peak position decreases to 5.24 Å. At 1600 K, the second peak split into sharp short and long peaks. The peaks increase in height maintains constant peak positions of r = 4.68 and 5.48 Å, respectively.

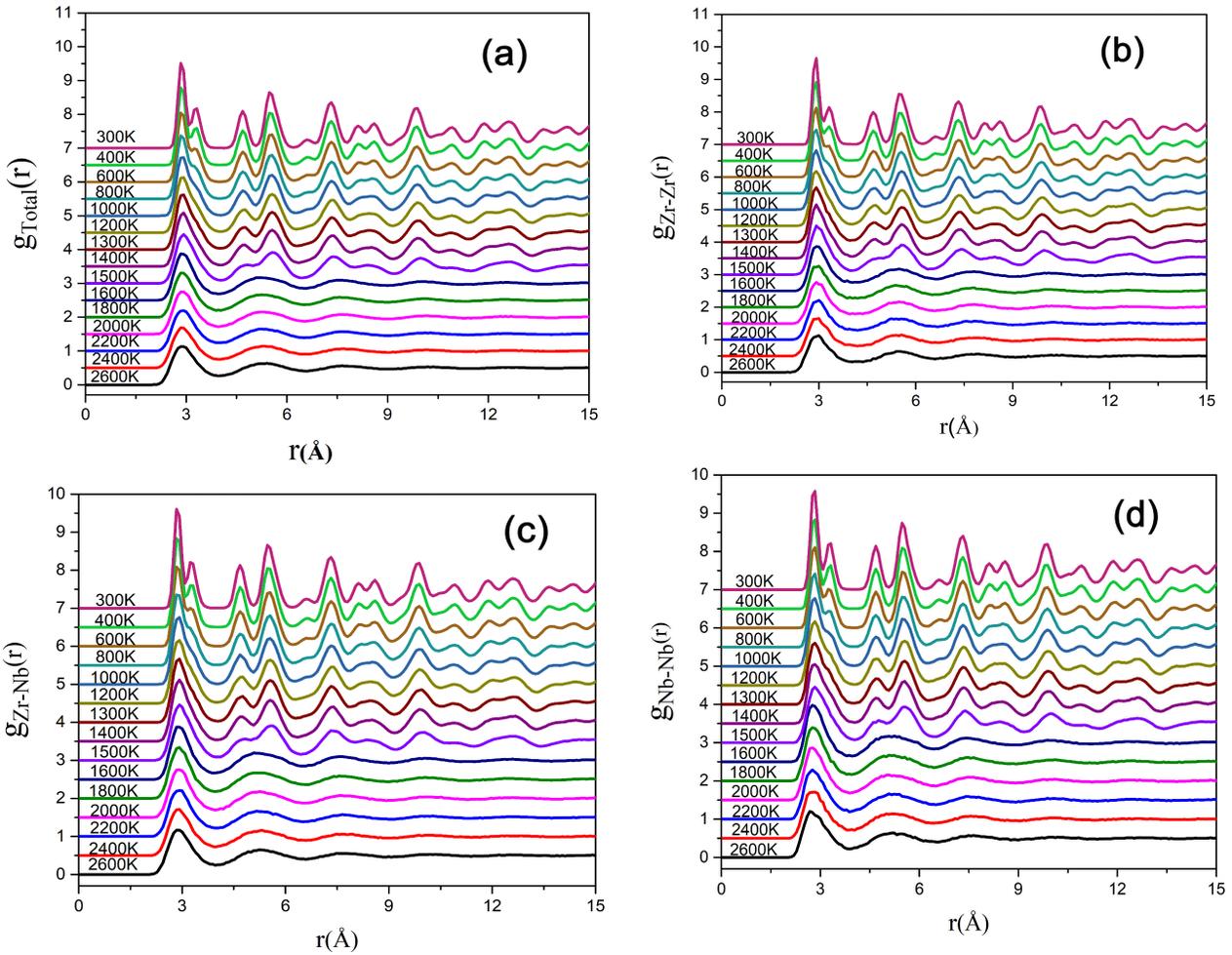

**Figure 7. (a)** Total PDFs for $Zr_{50}Nb_{50}$ under 15 Gpa pressure, from 2600 – 300 K, **(b), (c), (d)** PPDFs for Zr-Zr, Zr-Nb and Nb-Nb from 2600 – 300 K, respectively.

Zr-Nb Pairs show peak position fluctuation between 2.84 and 2.92 Å, from 2600 – 1500 K. Below 1500 K, the peak maxima shift to a constant value of 2.84 Å. At 600 K, the evolution of the shoulder peak is observed at r = 3.32 Å. This shoulder peak increases in height on further cooling from 600 – 300 K. The second peak maxima fluctuate between 5.16 Å and 5.24 Å from 2600 – 1700 K. At 1600 K the second peak splits into two short and long sharp peaks. These peaks increase in height on cooling down from 1500 – 300 K while maintaining a constant peak position at r = 4.68 and 5.48 Å, respectively.

The Nb-Nb pairs also show fluctuations in the first and second peak maxima position. At 2600 and 2400 K, the first peak maxima are located at r = 2.76 Å, and then increases to 2.84 Å at 1800 K. From 1700 – 300 K, the first peak maxima are always located at r = 2.76 Å except at 1600 K, 1000 K, and 400 K. At these temperature values it is located at r = 2.92, 2.84 and 2.84 Å respectively. The shoulder peak emerges and become obvious at 800 – 300 K. This shoulder peak is located at r = 3.24 Å. The second peak maxima also show fluctuation and splitting. At 2600 and 2400 K, the second peak is located at r = 5.16 Å and it fluctuates between 5.16 and 5.32 Å from, 2200 – 1700 K. At 1600 K the second peak splits into two short and long peaks located at r = 4.68 and 5.48 Å respectively. On further cooling down to 300 K, the peaks increase in height, while maintaining a constant peak position.

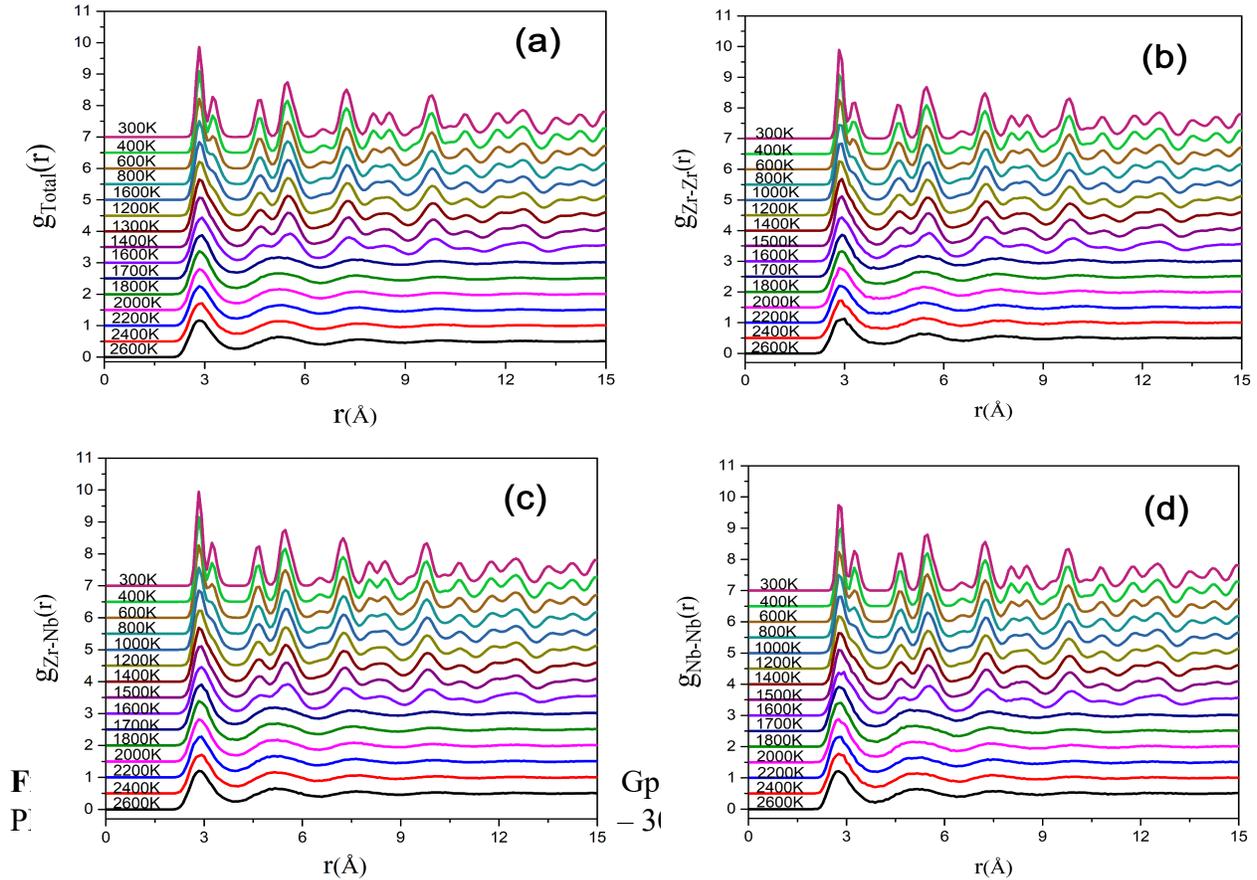

Zr-Zr pair's first peak maxima are located at r = 3 Å at 2600 K. However, on cooling from 2400 – 800 K, it is mostly located at r = 2.92 Å and maintains a constant value of 2.84 Å from, 800 – 300 K. At 600 K, the shoulder peak on the first main peak evolves at r = 3.24 Å, it increases in height
on further cooling to 300 K, at 300 K, this shoulder peak shifts to r = 3.32 Å. the second main peak mostly located at r = 5.4 Å from, 2600 – 1700 K. at 1600 K the second peak splits into two short and long peak' located at r = 4.68 Å and 5.56 Å. As we cool down from 1600 – 300 K, these two peaks increase in height while maintaining a constant peak position.

The snapshot of the atomic position under 0, 5, 10, 15, and 20 Gpa external pressure is shown in Figure 2 above. Under 0 and 5 Gpa external pressure the alloy shows randomly distributed atomic positions at high and low temperatures. Under 10, 15, and 20 Gpa external pressures the atomic positions, shows regular arrangement (long-range ordering) at 300 K.

### 3.3 Pressure and temperature dependency of coordination number

The coordination number (CN) is defined as the number of atoms bonded to a given atom in a structure and calculated using equation (3) below. To determine partial CN ($N_{ij}$) and total CN ($N_i$), the first minimum points of the PPDF and PDF have been taken as cut-off distances, respectively. By using PPDF, N of an element j in the first coordination shell of the element i is determined:

$$N_{ij} = \int 4\pi r \, \rho \, g\,(r)dr \qquad (4)$$

where $\rho_j$ is the density of the j atom, and rain is the position of the end of the first peak in $g_{ij}(r)$. Ni for atom i is calculated as $N_i = \Sigma_j N_{ij}$.

For $Zr_{50}Nb_{50}$ metallic melts, As the temperature decreases from 2600 K to 300 K, the changes in the total coordination numbers versus temperature are from 12.492 to 13.385, 13.02 to 13.5 under 0 Gpa and 5 Gpa external pressures respectively (Figure 9 (a), (b), respectively). The coordination number for Zr – centered increases from 12.56 to 13.8 and for Nb-centered it is from 12.492 to 12.97 from 2600 – 300 K. Moreover, the partial coordination number for each species is, also calculated. As temperature decreases from 2600 – 300 K, the Zr-Zr pairs show linear increases from 6.7 to 7.57 and 7.67 to 8.1 at 0 Gpa and 5 Gpa pressure respectively. The Nb-Nb pairs, also show a linear increase from 5.89 to 6.45 and 6.3 to 6.71 at 0 Gpa and 5 Gpa pressure, respectively. The Zr-Nb pairs always show small linear increases on cooling from 2600 – 300 K, it increases from 5.89 to 6.45 and 6.1 to 6.33 at 0 and 5 Gpa external pressure. The Coordination number shows an abrupt jump when external pressure increases to 10 Gpa, 15 Gpa, and 20 Gpa.

The total and partial coordination number jump is observed at 1400 K, 1500 K, and 1600 K, for 10, 5, 15, and 20 Gpa Figure 9 (c-e), external pressure respectively which coincides with E-T and V-T curve jump. The total coordination number increases linearly from 12.94 to 13.322 on quenching from 2600 to 1500 K, 13.3 to 13.5 on quenching from 2600 to 1600 K, and 13.59 to 13.67 on quenching from 2600 to 1800 K, under 10, 15 and 20 Gpa respectively. Jump is observed at 1400 K, 1500 K, and 1600 K, for respective pressure values. As one can see from the Figure, the total coordination number jumps from 13.322 at 1500 K to 13.75 at 1400 K, 13.5 at 1600 K to 13.91 at 1500 K, and 13.67 at 1800 K to 13.92 at 1600 K under 10, 15, and 20 Gpa external pressure. After an abrupt jump, the coordination number linearly increases to 14 for all pressure values.

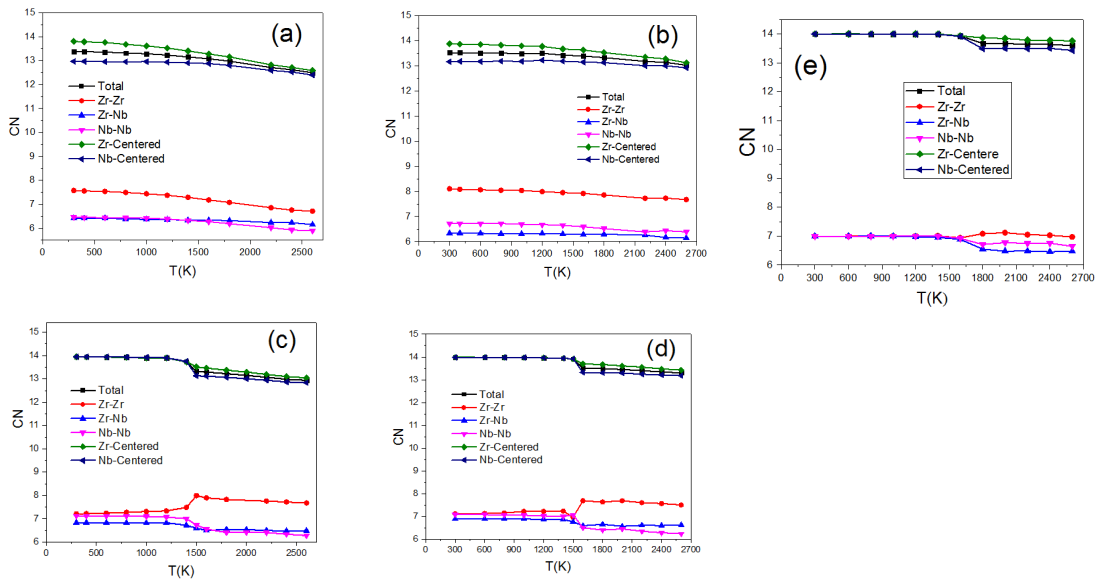

**Figure 9.** Evolution of Total and partial coordination number with temperature, under 0,5,10,15, and 20 Gpa external pressure., Insets (a-d), respectively.

The Zr-centered and Nb-centered as well as the partial coordination number are also calculated as shown in Fig.8. The Zr- centered coordination number, always shows the higher coordination number at all pressures. It increases linearly from 13.04 to 13.51 on quenching from 2600 – 1500 K, 13.42 to 13.69 on quenching from 2600 – 1600 K, and from 13.75 to 13.86 on quenching from 2600 – 1800 K under 10, 15, and 20 Gpa external pressure respectively. The Nb-centered coordination number also increases but is found below the Zr-centered coordination number. It increases linearly from 12.83 to 13.12 on quenching from 2600 – 1500 K, 13.17 to 13.31 on quenching from 2600 – 1600 K, and from 13.43 to 13.49 on quenching from 2600 – 1800 K

under 10, 15, and 20 Gpa external pressure. Under all these, respective pressure both Zr-centered and Nb -centered coordination shows sudden jump, after sudden jump all the coordination number increases slowly to 14 at 300 K.

The partial coordination number for Zr-Zr increases from 7.67 to 7.99 on cooling from 2600 - 1500 K. At 1400 K, Zr-Zr pairs also show an abrupt decrease to a value of 7.48 and linearly decreases to 7.12 from 1200-300 K under 10 Gpa external pressure. Also increases from 7.5 to 7.68 from 2600 – 1600 K, and 6.97 to 7.02 from 2600 -1700 K, under 15 and 20 Gpa external pressure respectively. It shows abrupt decreases to 7.06 at 1500 K and increases linearly to 7.12 from 1400 -300 K under 15 Gpa external pressure. Under 20 Gpa, the coordination number increases from 6.97 -7.02 from 2600 – 1700 K, and shows abrupt decreases to 6.94 at 1600 K, and slowly increases to 6.99 from 1400 – 300 K.

The partial coordination number for Nb-Nb under 10Gpa external pressure increases almost linearly from 6.27 to 6.4 on cooling from 2600 -1800 K. Abrupt increases are observed between 1800 and 1400 K. On further cooling from 1400 – 300 K, the coordination number linearly increases to 7.12. At 15Gpa the Nb-Nb partial coordination number increases from 6.24 to 6.5 from 2600 – 1600 K. Abrupt increase to 7.06 is observed at 1500 K, slowly increases to 7.08 on further cooling from 1400-300 K. Under 20 Gpa external pressure the Nb-Nb coordination number increases from 6.54 to 6.664 from 2600-1800 K followed by a decrease to 6.5 at 1700 K. Abrupt increase to 6.88 at 1600 K. Finally, linearly increases to 6.99 from 1400 – 300 K.

The Zr-Nb partial coordination number under 10 Gpa external pressure, shows slight increases from 6.48 to 6.54 on quenching from 2600 – 1800 K, followed by a decrease to 6.49 at 1600 K. Abrupt decrease between 1600 and 1400 K to 6.7 is observed followed by a slow increase to 6.83 from 1200 – 300 K. Under 15 Gpa external pressure it decreases from 6.63 to 6.57 at 2000 K, then increases to 6.65 followed by a decrease to 6.6 at 1500 K. Abrupt increase to 6.87, is observed between 1500 and 1400 K. It shows the almost constant value of 6.89 from 1400 – 300 K. Under 20 Gpa external pressure, it increases from 6.47 to 6.54 on quenching from 2600 to 1800 K. Abrupt increase is observed between 1800 and 1600 K. In this temperature region abrupt increase from 6.4 to 6.88 is detected. It linearly increases to 6.991 on further quenching from 1400 -300 K.

### 3.4 Bond angle distribution analysis

To further investigate the evolution of the SRO with temperature in $Zr_{50}Nb_{50}$ alloys, we perform the bond angle distribution function (shown in Figure 10.), $g_3(\theta)$, calculation, which can provide information about the local symmetry by measuring the statistical distribution of the bond-angle formed by three atoms within the first shell. The details of the formulation are given in reference [29]

The calculated bond angle distribution evolution versus temperature, under 0, 5, 10, 15, and 20 Gpa, external pressure is shown in Figure 10. (a-e). Here the first minima in the total g(r)s are used as a cutoff distance (4.04, 3.96, 4.12, 3.96, and 3.86 Å) are used for 0, 5,10,15,20 Gpa external pressures respectively. Under 0 and 5 Gpa external pressure the calculated bond angle distribution shows two prominent peaks near $\theta = 55.5°$ and $105.5°$ at 2600 K. On cooling down from 2600 – 1500 K, the first and second peaks shifts to higher angles, $\theta = 56.5°$ and $108.5°$ respectively. At 1200 K, a hump at $\theta = 146.5°$ and $\theta = 142.5°$ for 0 and 5 Gpa external pressure are observed respectively. On further cooling down to 300 K, under 0 Gpa the first peak increases to $\theta = 57.5°$ and the second peak increases to $\theta = 108.5°$, and the hump increase to $\theta = 148.5°$. While under 5 Gpa external pressure at 300 K, the first peak increases to $\theta = 57.5°$ and the second peak also increases to $110.5°$, and the hump increases to $149.5°$. The bond angle for the ideal icosahedra is ($\theta = 63.5°$ and $116.5°$) as shown in the figure (the orange line), on both pressures the peak positions are not far from the ideal icosahedra peak positions. The hump at $\theta = 148.5°$ indicates the complexity of the SRO in addition to the ISRO.

The Bond angle distribution for 10,15 and 20 Gpa, external pressure is also shown in Figure 10 inset (c-e). Under 10 Gpa external pressure, the first and second broad peaks are located at $\theta = 56.5°$ and the second peak at $\theta = 108°$ from 2600 – 1500 K. Below 1400 K, the peaks are evolved at $\theta = 70.5°$, $89.5°$, $109.5°$, and $124.5°$. They are in good agreement with the angle distribution of the BCC-like crystal ($\theta = 55°$, $\theta = 71°$, $90°$, $109°$, and $125°$)[29]. The melt under 15 and 20 Gpa, exactly shows a similar bond angle distribution besides, the peaks for BCC-like crystals occur below 1500 K and 1600 K respectively. Also, the first peak under 20 Gpa external pressure shifts from $\theta = 54.5°$ to $56.5°$.

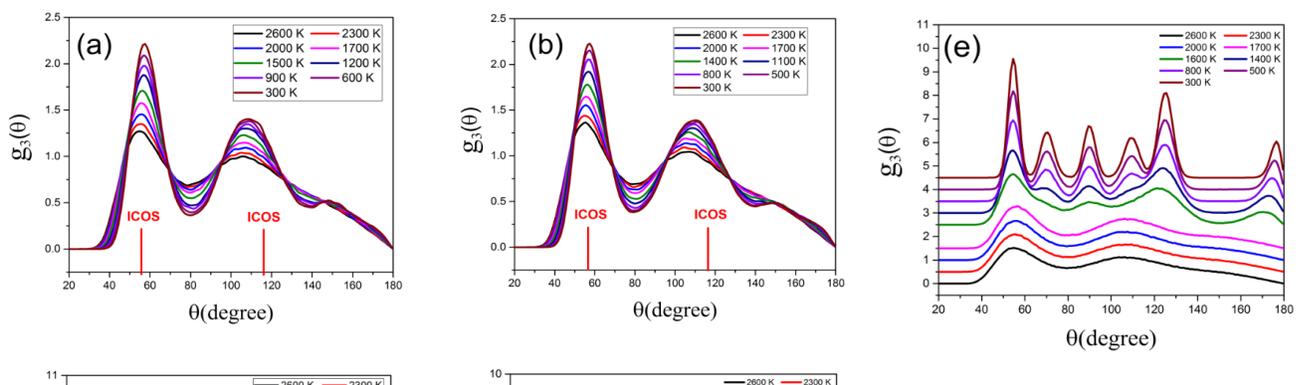

**Figure 10.** Evolution of bond angle distribution, versus temperature from 2600-300 K, under 0, 5, 10, 15, and 20 Gpa insets (a-e) respectively, (c-e) are shifted upward for clarity.

## 3.5 Voronoi tessellation analysis

To get a complete 3D geometrical view of the bonded cluster within the first coordination shell, we employ the Voronoi polyhedral analysis method [31,32]. Voronoi polyhedron (VP) reveals the atomic configuration of the central atom and its surrounding neighbor atoms. The Voronoi index $<n_3, n_4, n_5, n_6>$ is used to label the VP, where $n_i$ represents the number of i-edged faces on the surface of VP around the central atom, and the sum of all the faces ($\sum_i n_i$, ) of the voronoi polyhedral gives the coordination number of the local cluster.

The most abundant Voronoi polyhedral is presented in $Zr_{50}Nb_{50}$ alloy at different temperatures and pressures. were displayed in Figure 11. A lot of different polyhedrons are observed, some with a very small fraction. Thus, only the abundant ones are presented. Under 0 and 5 Gpa pressure, see Figure 11. (a and b), the system was dominated by 12 and 13 coordinated polyhedral. The fraction percent of perfect icosahedra <0,0,12,0> increases, from 1.5 to 7.5% on quenching from 1800 – 300 K under 0 Gpa and from 1.8 to 5.8% on quenching from 1200 – 300 K under 5 Gpa external pressure respectively. While the distorted icosahedra also increase on cooling from 2600 - 300 K. For instance, the indices with <0,1,10,2> and <0,2,8,4> increases from 0.9 to 8.5% and 1.4 to 6.5% under 0, 5 Gpa external pressure respectively. The distorted BCC-like cluster <0,3,6,4> was found at a larger fraction percent under 0 and 5 Gpa external pressures, at higher and lower temperatures. It increases from 2.2 to 9.1% on cooling from 2600 – 300 K and from 2.7 to 10.2% under 0 and 5 Gpa external pressure.

The fraction percent of icosahedra both distorted and perfect is reduced significantly below 1400 and 1500 K under 10 and 15 Gpa, see Figure 11 (c and d) external pressure respectively. For instance, the fraction percent of indices of perfect icosahedra <0,0,12,0> was only observed with vea very low fraction (1.4%) at 1400 K under 10 Gpa external pressure and at 1400 K under 15 Gpa external pressure.

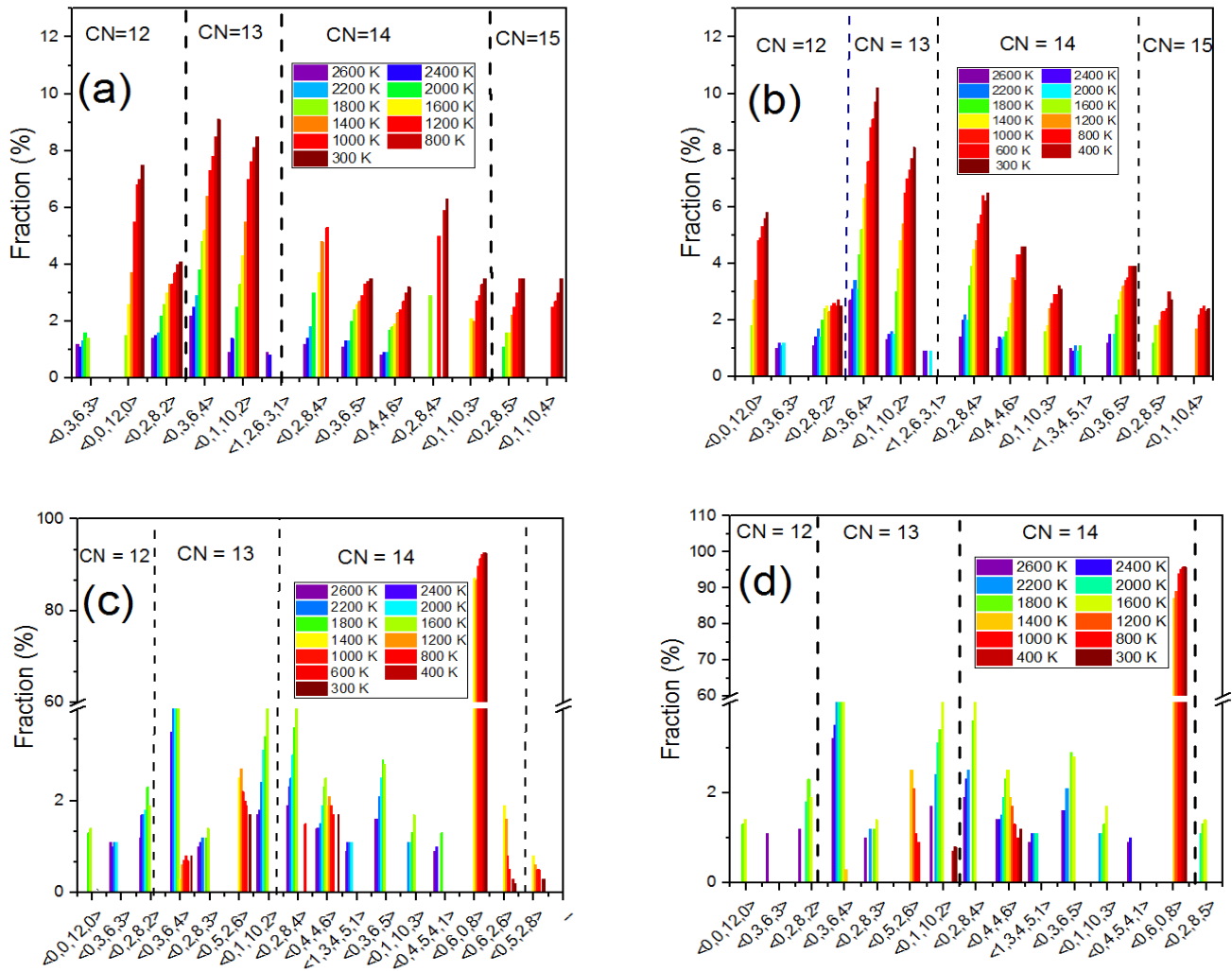

**Figure 11**. Evolution of voronoi index from 2600 – 300 K, under external pressure 0, 5, 10, and 15 Gpa insets (a-d) respectively.

Under 10 Gpa external pressure, the distorted icosahedra, indices <0,2,8,2>, <0,1,10,2> and <0,2,8,4> increases on cooling from 2600 – 1500 K and disappears below 1400 K. The distorted BCC-like cluster <0,3,6,4> and increases from 3.5 to 5.3% on cooling from 2600 – 1400 K, and it decreases to 0.8% at 300 K. Also, the other BCC-like cluster <0,4,4,6> increases from 1.4 to 2.5% on cooling from 2600- 1400 K, and decreases to 1.7 % below 1400 K. The fraction percent of BCC-like cluster <0,6,0,8> dominates the whole system below 1400 K, it increases from 87.1 to 92.6% from 1400 - 300 K.

Under 15 Gpa external pressure, the fraction the distorted icosahedra, <0,1,10,2> and <0,2,8,4> increase from 1.7 to 4.2% and 1.9 to 4.2% on cooling from 2600 – 1500 K respectively and they disappear below 1500 K. While the mixed cluster type <0,3,6,5> also increases from 1.6 to 2.8% on cooling from 2600 – 1500 K and disappears below 1500 K. The distorted BCC-like cluster <0,3,6,4> and <0,4,4,6> shows an increase from 3.2 to 5.3% and 1.4 to 2.5% on cooling from 2600 – 1500 K. Below 1500 K, the fraction percent of <0,6,0,8> BCC-like cluster dominates the system, and comprises the fraction percent of 95.8% at 300 K.

## 4 Conclusions

The solidification of a $Zr_{50}Nb_{50}$ melt in a temperature range of 4000 – 300 K under 0, 5, 10, 15, and 20 Gpa external pressure was studied using a classical molecular dynamics simulation. A cooling rate of $10^{12}$K/s was used. Using the radial distribution function g(r), coordination number, bond angle distribution, and Voronoi mosaic analysis method, we found that the icosahedral motifs (which are the signature of short-range order) and distorted BCC-like clusters dominate in the liquid and glass region under 0 and 5 Gpa external pressure. The phase transition to the crystalline structure was observed at 10, 15, and 20 Gpa external pressure. At these pressures above the glass transition temperatures, the system consisted of icosahedral and distorted BCC-like clusters. Below the glass transition temperature, all analyzes showed that $Zr_{50}Nb_{50}$ crystallized to a BCC-like structure. Furthermore, increasing pressure results in a significant increase in crystal-like clusters, implying that pressure may induce a structural transition in the systems from a more amorphous to a more crystal-like structure.

# Acknowledgments

This work was supported by a thematic research project (Grant no. TR/036/2020) funded by Addis Ababa University. Moreover, this research used the computing facility of the Center for Functional Nanomaterials (CFN), which is a U.S. Department of Energy Office of Science User Facility, at Brookhaven National Laboratory under Contract No. DE-SC0012704 and project number 308379.

# References

[1]     F. F. Abraham, "Comment on 'An isothermal-isobaric computer simulation of the supercooled-liquid/glass transition region: Is the short-range order in the amorphous solid fee?,'" *J. Chem. Phys.*, vol. 75, no. 1, pp. 498–499, 1981, doi: 10.1063/1.441852.

[2]     D. B. Miracle, "A structural model for metallic glasses," *Nat. Mater.*, vol. 3, no. 10, pp. 697–702, 2004, doi: 10.1038/nmat1219.

[3]     X. J. Liu *et al.*, "Atomic packing symmetry in the metallic liquid and glass states," *Acta Mater.*, vol. 59, no. 16, pp. 6480–6488, 2011, doi: 10.1016/j.actamat.2011.07.012.

[4]     H. W. Sheng, W. K. Luo, F. M. Alamgir, J. M. Bai, and E. Ma, "Atomic packing and short-to-medium-range order in metallic glasses," *Nature*, vol. 439, no. 7075, pp. 419–425, 2006, doi: 10.1038/nature04421.

[5]     X. W. Fang *et al.*, "Spatially resolved distribution function and the medium-range order in metallic liquid and glass," *Sci. Rep.*, vol. 1, pp. 19–21, 2011, doi: 10.1038/srep00194.

[6]     L. Ward, D. Miracle, W. Windl, O. N. Senkov, and K. Flores, "Structural evolution and kinetics in Cu-Zr metallic liquids from molecular dynamics simulations," *Phys. Rev. B - Condens. Matter Mater. Phys.*, vol. 88, no. 13, pp. 1–10, 2013, doi: 10.1103/PhysRevB.88.134205.

[7]     A. Inoue, B. Shen, H. Koshiba, H. Kato, and A. R. Yavari, "Cobalt-based bulk glassy alloy with ultrahigh strength and soft magnetic properties," *Nat. Mater.*, vol. 2, no. 10, pp. 661–663, 2003, doi: 10.1038/nmat982.

[8]     M. D. Demetriou *et al.*, "A damage-tolerant glass," *Nat. Mater.*, vol. 10, no. 2, pp. 123–128, 2011, doi: 10.1038/nmat2930.

[9] L. Tian *et al.*, "Approaching the ideal elastic limit of metallic glasses," *Nat. Commun.*, vol. 3, pp. 1–6, 2012, doi: 10.1038/ncomms1619.

[10] Y. H. Jeong, K. Ok, and H. G. Kim, "Correlation between microstructure and corrosion behavior of Zr – Nb binary alloy," vol. 302, pp. 9–19, 2002.

[11] M. Z. Ma *et al.*, "Wear resistance of Zr-based bulk metallic glass applied in bearing rollers," *Mater. Sci. Eng. A*, vol. 386, no. 1–2, pp. 326–330, 2004, doi: 10.1016/j.msea.2004.07.054.

[12] Y. Q. Cheng and E. Ma, "Atomic-level structure and structure-property relationship in metallic glasses," *Prog. Mater. Sci.*, vol. 56, no. 4, pp. 379–473, 2011, doi: 10.1016/j.pmatsci.2010.12.002.

[13] D. Holland-Moritz, S. Stüber, H. Hartmann, T. Unruh, T. Hansen, and A. Meyer, "Structure and dynamics of liquid Ni36 Zr64 studied by neutron scattering," *Phys. Rev. B - Condens. Matter Mater. Phys.*, vol. 79, no. 6, pp. 1–8, 2009, doi: 10.1103/PhysRevB.79.064204.

[14] T. T. Debela *et al.*, "Atomic structure evolution during solidification of liquid niobium from ab initio molecular dynamics simulations," *J. Phys. Condens. Matter*, vol. 26, no. 5, 2014, doi: 10.1088/0953-8984/26/5/055004.

[15] T. T. Debela, X. D. Wang, Q. P. Cao, D. X. Zhang, J. J. Zhu, and J. Z. Jiang, "Phase selection during solidification of liquid magnesium via ab initio molecular dynamics simulations," *J. Appl. Phys.*, vol. 117, no. 11, pp. 1–8, 2015, doi: 10.1063/1.4914414.

[16] L. Zhou, J. Pan, L. Lang, Z. Tian, Y. Mo, and K. Dong, "Atomic structure evolutions and mechanisms of the crystallization pathway of liquid Al during rapid cooling," *RSC Adv.*, vol. 11, no. 63, pp. 39829–39837, 2021, doi: 10.1039/d1ra06777j.

[17] E. D. Zanotto and J. C. Mauro, "The glassy state of matter: Its definition and ultimate fate," *J. Non. Cryst. Solids*, vol. 471, no. March, pp. 490–495, 2017, doi: 10.1016/j.jnoncrysol.2017.05.019.

[18] Y. Li, S. Zhao, Y. Liu, P. Gong, and J. Schroers, "How Many Bulk Metallic Glasses Are There?," *ACS Comb. Sci.*, vol. 19, no. 11, pp. 687–693, 2017, doi: 10.1021/acscombsci.7b00048.

[19] V. O. Kharchenko and D. O. Kharchenko, "Ab-initio calculations for structural properties of Zr-Nb alloys," *Condens. Matter Phys.*, vol. 16, no. 1, pp. 1–8, 2013, doi: 10.5488/CMP.16.13801.

[20] S. J. Yang, L. Hu, L. Wang, and B. Wei, "Molecular dynamics simulation of liquid

structure for undercooled Zr-Nb alloys assisted with electrostatic levitation experiments," *Chem. Phys. Lett.*, vol. 701, pp. 109–114, 2018, doi: 10.1016/j.cplett.2018.04.050.

[21] D. E. Smirnova and S. V. Starikov, "An interatomic potential for simulation of Zr-Nb system," *Comput. Mater. Sci.*, vol. 129, pp. 259–272, 2017, doi: 10.1016/j.commatsci.2016.12.016.

[22] M. Cottura and E. Clouet, "Solubility in Zr-Nb alloys from first-principles," *Acta Mater.*, vol. 144, pp. 21–30, 2018, doi: 10.1016/j.actamat.2017.10.035.

[23] X. Wang, L. B. Liu, M. F. Wang, X. Shi, G. X. Huang, and L. G. Zhang, "Computational modeling of elastic constants as a function of temperature and composition in Zr-Nb alloys," *Calphad Comput. Coupling Phase Diagrams Thermochem.*, vol. 48, pp. 89–94, 2015, doi: 10.1016/j.calphad.2014.11.003.

[24] H. Dyja, A. Kawałek, and K. Ozhmegov, "Experimental studies on Zr–1%Nb alloy properties in technological conditions of cold pilger tube rolling process," *Arch. Civ. Mech. Eng.*, vol. 19, no. 1, pp. 268–273, 2019, doi: 10.1016/j.acme.2018.10.002.

[25] Y. H. Jeong, H. G. Kim, and T. H. Kim, "Effect of β phase, precipitate and Nb-concentration in matrix on corrosion and oxide characteristics of Zr-xNb alloys," *J. Nucl. Mater.*, vol. 317, no. 1, pp. 1–12, 2003, doi: 10.1016/S0022-3115(02)01676-8.

[26] S. S. Kliavinek and L. N. Kolotova, "Molecular Dynamics Simulation of Glass Transition of the Supercooled Zr–Nb Melt," *J. Exp. Theor. Phys.*, vol. 131, no. 2, pp. 284–297, 2020, doi: 10.1134/S1063776120080105.

[27] S. Y. Wu, S. H. Wei, G. Q. Guo, J. G. Wang, and L. Yang, "Structural mechanism of the enhanced glass-forming ability in multicomponent alloys with positive heat of mixing," *Sci. Rep.*, vol. 6, no. November, pp. 1–12, 2016, doi: 10.1038/srep38098.

[28] S. Sengul, M. Celtek, and U. Domekeli, "The structural evolution and abnormal bonding ways of the Zr80Pt20 metallic liquid during rapid solidification under high pressure," *Comput. Mater. Sci.*, vol. 172, no. September 2019, p. 109327, 2020, doi: 10.1016/j.commatsci.2019.109327.

[29] J. Hafner, "Bond-Angle Distribution Functions in Metallic Glasses," Le J. Phys. Colloq., vol. 46, no. C9, pp. C9-69-C9-78, 1985, doi: 10.1051/jphyscol:1985908.

[30] N. Jakse, O. Le Bacq, and A. Pasturel, "Prediction of the local structure of liquid and supercooled tantalum," *Phys. Rev. B - Condens. Matter Mater. Phys.*, vol. 70, no. 17, pp. 1–6, 2004, doi: 10.1103/PhysRevB.70.174203.

[31] V. A. Borodin, "Local atomic arrangements in polytetrahedral materials," *Philos. Mag. A*

*Phys. Condens. Matter, Struct. Defects Mech. Prop.*, vol. 79, no. 8, pp. 1887–1907, 1999, doi: 10.1080/01418619908210398

[32] J. L. Finney and P. R. S. L. A, "Random packings and the structure of simple liquids. I. The geometry of random close packing," Proc. R. Soc. London. A. Math. Phys. Sci., vol. 319, no. 1539, pp. 479–493, 1970, doi: 10.1098/rspa.1970.0189